\documentclass[twocolumn, pdflatex]{aastex631}

\usepackage{amsmath}
\usepackage{color}
\usepackage{url}
\newcommand{\red}[1]{\textcolor{red}{#1}}

\usepackage{subfiles} 

\begin{document}

\title{A self-consistent model of shock-heated plasma in non-equilibrium states for direct parameter constraints from X-ray observations}

\correspondingauthor{Yuken Ohshiro}
\email{ohshiro-yuken@g.ecc.u-tokyo.ac.jp}

\author{Yuken Ohshiro}
\affiliation{Department of Physics, Graduate School of Science, The University of Tokyo, 7-3-1 Hongo, Bunkyo-ku, Tokyo 113-0033, Japan}
\affiliation{Institute of Space and Astronautical Science (ISAS), Japan Aerospace Exploration Agency (JAXA), 3-1-1 Yoshinodai, Chuo-ku, Sagamihara, Kanagawa 252-5210, Japan}

\author{Shunsuke Suzuki}
\affiliation{Department of Science and Engineering, Graduate School of Science and Engineering, Aoyama Gakuin University, 5-10-1 Fuchinobe, Sagamihara, Kanagawa 252-5258, Japan}
\affiliation{Institute of Space and Astronautical Science (ISAS), Japan Aerospace Exploration Agency (JAXA), 3-1-1 Yoshinodai, Chuo-ku, Sagamihara, Kanagawa 252-5210, Japan}

\author{Yoshizumi Okada}
\affiliation{Department of Science and Engineering, Graduate School of Science and Engineering, Aoyama Gakuin University, 5-10-1 Fuchinobe, Sagamihara, Kanagawa 252-5258, Japan}
\affiliation{Institute of Space and Astronautical Science (ISAS), Japan Aerospace Exploration Agency (JAXA), 3-1-1 Yoshinodai, Chuo-ku, Sagamihara, Kanagawa 252-5210, Japan}

\author{Hiromasa Suzuki}
\affiliation{Institute of Space and Astronautical Science (ISAS), Japan Aerospace Exploration Agency (JAXA), 3-1-1 Yoshinodai, Chuo-ku, Sagamihara, Kanagawa 252-5210, Japan}

\author{Hiroya Yamaguchi}
\affiliation{Institute of Space and Astronautical Science (ISAS), Japan Aerospace Exploration Agency (JAXA), 3-1-1 Yoshinodai, Chuo-ku, Sagamihara, Kanagawa 252-5210, Japan}
\affiliation{Department of Physics, Graduate School of Science, The University of Tokyo, 7-3-1 Hongo, Bunkyo-ku, Tokyo 113-0033, Japan}



\begin{abstract}
    X-ray observations of shock-heated plasmas, such as those found in supernova remnants, often exhibit features of temperature and ionization non-equilibrium.
    For accurate interpretation of these observations, proper calculations of the equilibration processes are essential. 
    Here, we present a self-consistent model of thermal X-ray emission from shock-heated plasmas that accounts for both temperature and ionization non-equilibrium conditions.
    For a given pair of shock velocity and initial electron-to-ion temperature ratio, the temporal evolution of the temperature and ionization state of each element was calculated by simultaneously solving the relaxation processes of temperature and ionization.
    The resulting thermal X-ray spectrum was synthesized by combining our model with the AtomDB spectral code. 
    Comparison between our model and the \texttt{nei} model, a constant-temperature non-equilibrium ionization model available in the XSPEC software package, reveals a 30\% underestimation of the ionization timescale in the \texttt{nei} model.
    We implemented our model in XSPEC to directly constrain the shock wave properties, such as the shock velocity and collisionless electron heating efficiency, from the thermal X-ray emission from postshock plasmas.
    We applied this model to archival Chandra data of the supernova remnant N132D, providing a constraint on the shock velocity of $\sim 800~\mathrm{km\,s^{-1}}$, in agreement with previous optical studies.
    \end{abstract}



\section{Introduction} \label{sec: intro}

Astrophysical shock waves are ubiquitously observed in various objects in different spatial scales, 
such as solar wind \citep[e.g.,][]{reames1999}, 
gamma-ray bursts \citep[e.g.,][]{paradijs2000}, 
and supernovae \citep[e.g.,][]{tatischeff2018}.
In those astrophysical plasmas, the number density of particles is so low ($\lesssim 1\,\mathrm{cm^{-3}}$) that the mean free path of Coulomb interactions is much longer than the thickness of the shock transition region.
Therefore, collisionless processes involving collective interactions between particles and electromagnetic fields are thought to play an important role in the shock transition \citep[e.g.,][]{heng2010,ghavamian2013}. 
Supernova remnants (SNRs) offer an ideal site to study such collisionless shocks because their optical or X-ray emission is relatively bright and well resolved spatially due to the large spatial extent and proximity 
\citep[e.g.,][]{vink2003, rakowski2005}.

Because of the inefficient energy transfer between different species 
at the collisionless limit, it is expected that the post-shock temperature of each species depends on the particle mass as 

\noindent
\begin{gather}
    \label{eq: shock heating}
    kT_{i} = \frac{3}{16} m_{i} {V_{\mathrm{sh}}}^{2},
\end{gather}
\noindent
where $k$, $m_{i}$ and $V_{\mathrm{sh}}$ are the Boltzmann constant,  mass of species $i$, and shock velocity, respectively.
In principle, Equation~\ref{eq: shock heating} is valid for electrons as well.
Therefore, the electron temperature is expected to be three orders of magnitude lower than the ion temperature in the regions immediately behind the collisionless shock.
Some observations of post-shock plasmas in SNRs indeed showed that the ion temperatures are mass-proportional \citep{raymond2017, miceli2019}.
However, several observations and theoretical investigations have suggested that an instantaneous energy transport from ions to electrons takes place at the shock front (so-called ``collisionless electron heating''), so the downstream electron temperature becomes substantially higher than the prediction of Equation~\ref{eq: shock heating} \citep[e.g.,][]{rakowski2005,broersen2013,yamaguchi2014a,miceli2023, raymond2023}.
Notably, an optical study of young and middle-aged SNRs indicates that the post-shock electron-to-proton temperature ratio is inversely proportional to the square of the shock velocity, implying that the shock velocity is closely related to the efficiency of collisionless electron heating \citep{ghavamian2007}.

The non-equilibrium state formed behind a collisionless shock is then gradually relaxed via Coulomb interactions in further downstream regions. 
In fact, an increase in electron temperature due to this process was recently observed in Tycho's SNR \citep{matsuda2022}.
Once electrons gain adequate thermal energy, they can contribute to the ionization of heavy elements via collisions.
The timescale for collisional plasma to reach the ionization equilibrium is comparable to that of the temperature equilibrium, both of which are $\sim$\,$10^{12}$\,$(n/1\,{\rm cm}^{-3})^{-1}$\,s 
\citep[e.g.,][]{Vink2020,yamaguchi2022}. 
Evidence for non-equilibrium ionization (NEI) is commonly observed in young SNRs with ages less than $\sim$\,$10^4$\,yr \citep[e.g.,][]{Yamaguchi2014}. 
Therefore, proper treatment of the equilibration processes is crucial in interpreting observed spectra of SNRs, which in turn provides the key to understanding both shock physics \citep[e.g.,][]{vink2003, yamaguchi2014a} and supernova progenitors \citep[e.g.,][]{badenes2006, Ohshiro2021}.

Theoretical models of X-ray emission from non-equilibrium plasmas have continuously been developed in previous work \citep[e.g.,][]{hamilton1983, borkowski2001} and 
extended for different purposes: e.g., to constrain the progenitors, explosion mechanisms, and surrounding environments
\citep{badenes2006, orlando2015, kobashi2024}, 
to explore the mechanisms cosmic-ray acceleration at the shock front \citep{ellison2007, lee2012, ferrand2014, shimoda2022}, 
and to constrain the efficiency of the collisionless electron heating and the kinematics of SNRs
\citep{miceli2019, sapienza2024}.
However, most of these models were developed to interpret observations of specific SNRs and are inapplicable directly to other SNRs in general.

To date, publicly-available spectral models, such as \texttt{nei}\footnote{\url{https://heasarc.gsfc.nasa.gov/xanadu/xspec/manual/node195.html}} and \texttt{pshock}\footnote{\url{https://heasarc.gsfc.nasa.gov/xanadu/xspec/manual/node217.html}} in the XSPEC software package\footnote{\url{https://heasarc.gsfc.nasa.gov/xanadu/xspec/}},
have been frequently used to analyze X-ray spectra of thermal SNRs. 
Although these models calculate the ionization balance in the NEI conditions, the temperature equilibration processes are not taken into account, and instead, a constant electron temperature is assumed throughout the time evolution of the NEI state. 
Since the ionization and recombination rates as well as the line emissivities depend on the electron temperature, the current models may introduce some biases in the estimation of the ionization timescale (i.e., a product of the electron density and time elapsed after the plasma is shock-heated). This potentially affects the interpretation of the observations of SNRs, especially with high-resolution X-ray spectroscopy enabled by the XRISM satellite \citep{XRISM2020}.

In this paper, we construct a self-consistent model for shock-heated plasmas that directly relates the physical quantities related to the shock heating (i.e.,  the shock velocity and electron-ion temperature ratio at the immediate post-shock region) to observed X-ray spectra.
We also investigate the systematic biases in plasma parameters introduced by traditional NEI models.
The methodologies and assumptions of our model are described in Section \ref{sec: method}.
In Section \ref{sec: results}, we calculate the temporal evolution of the post-shock plasma and resulting X-ray spectra for two different cases in terms of elemental composition, the solar-abundance interstellar medium (ISM) and metal-rich ejecta. 
In Section \ref{sec: discussion}, systematic biases in the plasma parameters (e.g., ionization timescale and elemental abundances) introduced by the \texttt{nei} model are evaluated and discussed.
We also discuss the time evolution of ion temperatures and thermal Doppler broadening that is measurable with XRISM.
In Section~\ref{sec: Application}, we apply our model to archival Chandra observations of the SNR N132D to demonstrate that the shock velocity can directly be constrained from a fit to its X-ray spectra.
Finally, we conclude this study in Section~6.

The XSPEC-readable model files used in this work are available on our website (TBA).

\section{Method} \label{sec: method}

In this section, we construct a framework to model the thermal evolution of shock-heated plasmas and their X-ray emission by calculating the temperature and ionization equilibration processes simultaneously.
We consider a plane-parallel shock wave propagating into a uniform-density medium.
Our model does not take into account cosmic-ray acceleration and other non-adiabatic processes, such as radiative cooling and thermal conduction into the ambient medium.
Therefore, the scope of our model is limited to thermal plasmas in postshock regions with no significant effect of these processes.
Incorporation of these effects will be addressed in future studies.

As the initial conditions, we assume that all atoms are singly ionized regardless of the collisionless electron heating efficiency.
The initial temperature of each species is determined by Equation~\ref{eq: shock heating} when the collisionless electron heating does not work at all.

\begin{figure}[t]
    \centering
    \includegraphics[width=8.5cm]{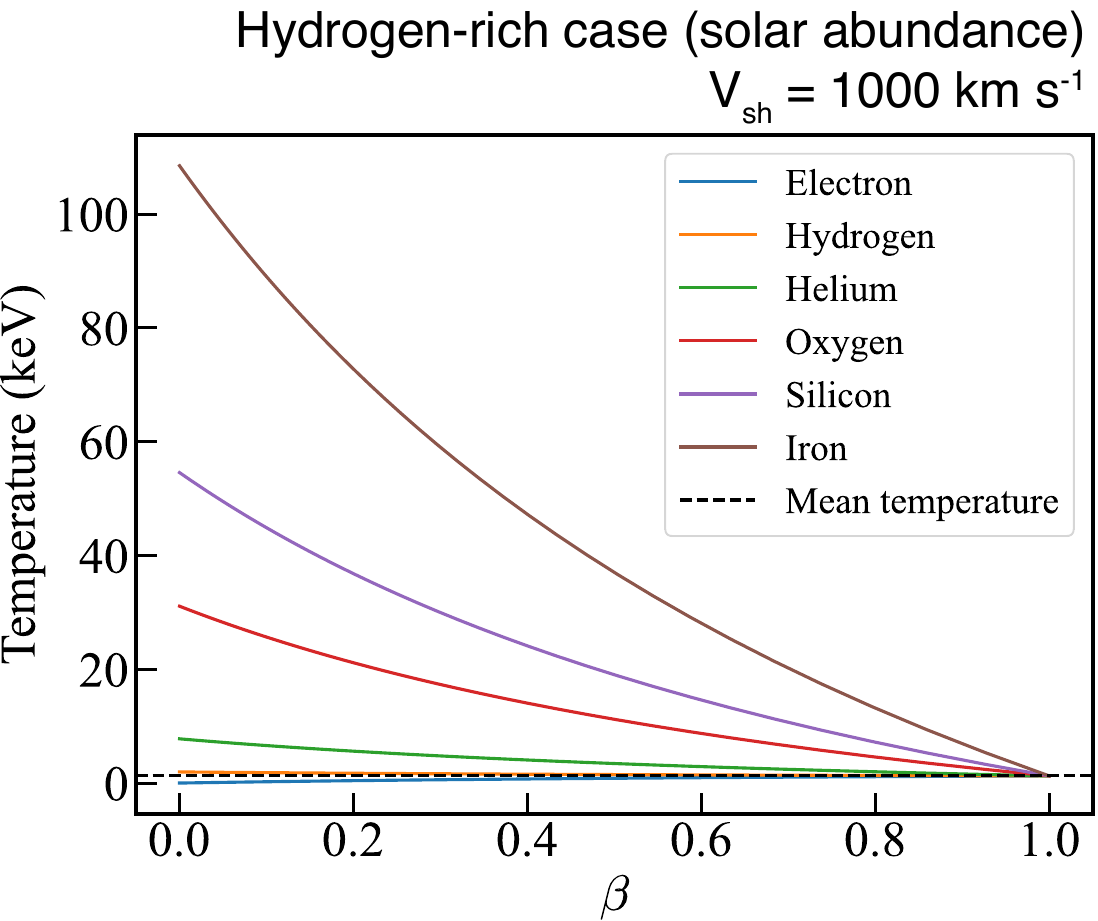}
    \caption
    {
        Immediate post-shock temperatures of each particle species as a function of $\beta$ in a solar abundance plasma heated by a shock wave with a velocity of $V_{\mathrm{sh}} = 1000\,\mathrm{km\,s^{-1}}$.
    }
    \label{fig: CollisionlessShockEffect}
\end{figure}

The total internal energy density $\epsilon_\mathrm{tot}$ of the shocked plasma composed of an ideal gas is then given as: 
\noindent
\begin{gather}
    \label{eq: total}
    \epsilon_{\mathrm{tot}} = \sum_i \epsilon_{\mathrm{i}} = \sum_i \frac{3}{2} n_i kT_i
\end{gather}
\noindent
where $n_i$ is the number density of species $i$.
Next, we introduce the effect of collisionless electron heating by parameterizing the electron-to-ion temperature ratio as $\beta = kT_{\mathrm{e}}/kT_{\mathrm{ion}}$, where $kT_{\mathrm{ion}} = \sum_X n_X kT_{\mathrm{X}}/ \sum_X n_X$ is the mean ion temperature.
The internal energy density ratio between electrons and ions after considering the collisionless electron heating is, therefore, obtained as follows:
\noindent
\begin{gather}
    \label{eq: internal energy ratio}
    \frac{\epsilon_{\mathrm{e}}}{\epsilon_{\mathrm{ion}}}
    = \frac{\frac{3}{2} n_{\mathrm{e}} kT_{\mathrm{e}}}{\sum_X \frac{3}{2} n_{\mathrm{X}} kT_{\mathrm{X}}}
    = \frac{kT_\mathrm{e}}{kT_{\mathrm{ion}}} \frac{n_{\mathrm{e}}}{\sum_X n_X}
    = \frac{\beta n_\mathrm{e}}{\sum_X n_X}.
\end{gather}
\noindent
We assume that the internal energy density given by Equation~\ref{eq: total} is conserved during the collisionless electron heating process.
Therefore, the internal energy density of electrons and ions after the collisionless electron heating is given by:
\begin{gather}
    \label{eq: each internal energy}
    \begin{aligned}
    \epsilon_{\mathrm{e}} &= \frac{\frac{\beta n_\mathrm{e}}{\sum_X n_X}}{1 + \frac{\beta n_\mathrm{e}}{\sum_X n_X}} \epsilon_{\mathrm{tot}}, \\
    \epsilon_{\mathrm{ion}} &= \frac{1}{1 + \frac{\beta n_\mathrm{e}}{\sum_X n_X}} \epsilon_{\mathrm{tot}}
    \end{aligned}
\end{gather}
\noindent
To determine the post-shock temperature of each particle species, additional conditions that constrain the temperature relationship need to be considered.
Here we introduce a simplified condition as follows:
\begin{gather}
    \label{eq: addtitional condition}
    \frac{kT_i - (kT_i)^{\prime}}{kT_i - kT} = \frac{kT_j - (kT_j)^{\prime}}{kT_j - kT},
\end{gather}
\noindent
Where $kT_i$ is the postshock temperature of species $i$ without the effect of collisionless electron heating (i.e., given in Equation~\ref{eq: shock heating}), $(kT_i)^{\prime}$ is the postshock temperature of species $i$ with the effect of collisionless electron heating, and $kT = \sum_i n_i kT_i/ \sum_i n_i$ is the density-weighted mean temperature of the postshock plasma.
Under these conditions, the post-shock temperature of each particle species is given by:
\begin{gather}
    \label{eq: postshock temperature}
    (kT_i)^{\prime} = \frac{\sum_i n_i kT_i + \sum_{j \neq i}[1 + \frac{\beta n_{\mathrm{e}}}{\sum_k n_k}] \frac{kT_j - kT_i}{kT_i - kT} n_j kT}{\sum_j [1 + \frac{\beta n_{\mathrm{e}}}{\sum_k n_k}] \frac{kT_j - kT}{kT_i - kT} n_j}
\end{gather}
\noindent
We show in Figure~\ref{fig: CollisionlessShockEffect} the post-shock temperatures of each species as a function of $\beta$ in case of $V_{\mathrm{sh}} = 1000\,\mathrm{km\,s^{-1}}$.

The allowable range of the $\beta$ value is  
$m_{\mathrm{e}}/\mu m_{\mathrm{p}} \leq \beta \leq 1$, 
where $\mu$ is the mean atomic weight of the ions (excluding electrons). 
The lowest value of $\beta$ (hereafter $\beta_{\rm min}$) corresponds to the case of no collisionless electron heating, whereas $\beta = 1$ indicates that the temperature equilibrium is immediately achieved in the post-shock plasma.

Thermal energy transfer via the Coulomb interactions between different species takes place in the downstream region.
It is assumed that a Maxwellian energy distribution is immediately achieved for each species at each time step.
While this assumption may not be exactory hold immediately after the collisionless shock front, but it is suggested to be sufficiently valid by the results of particle-in-cell simulations \citep{kropotina2016}.
The temperature change of species $i$ can be described as
\noindent
\begin{gather}
    \label{eq: temperature equilibrium}
    \frac{\mathrm{d}\epsilon_i}{\mathrm{d}t} = \sum_{j} \frac{\frac{n_i}{n_j} \epsilon_j - \epsilon_i}{t_{\mathrm{eq},\,ij}},
\end{gather}
\noindent
\citep{spitzer1956}.
$t_{\mathrm{eq},\,ij}$ is the timescale of temperature equilibrium between the species $i$ and $j$ given by
\noindent
\begin{gather}
    \label{eq: t_eq}
    t_{\mathrm{eq},\,ij} = \frac{3 m_i m_j}{8 (2\pi)^{1/2} n_j \bar{Z}^{2}_i \bar{Z}^{2}_j e^{4} \ln{\Lambda}} \left(\frac{kT_i}{m_i} + \frac{kT_j}{m_j} \right)^{3/2},
\end{gather}
\noindent
where $n_i$ and $\bar{Z}_i$ are the number density and mean charge number of the species $i$, respectively.
$\ln{\Lambda}$ is the Coulomb logarithm, and $\Lambda$ is given by
\noindent
\begin{gather}
    \Lambda = \frac{3}{2 \bar{Z} e^{3}} \left(\frac{k^{3}T^{3}_{\mathrm{e}}}{\pi n_\mathrm{e}}\right)^{1/2},
\end{gather}
\noindent
where $\bar{Z}$ and $n_{\mathrm{e}}$ are the mean charge number of all ions and the electron density of the plasma, respectively.

Ionization equilibrium can be described by the rate equation, representing the balance between ionization and recombination from each charge state.
In this work, we adopt a general formulation of the rate equation for the atomic number $X$ element:
\noindent
\begin{gather}
    \begin{split}
        \label{eq: rate equation}
        \frac{\mathrm{d}f_{X, x}}{\mathrm{d}t} &=
        n_{\mathrm{e}}[S_{X, x-1}(T_{\mathrm{e}})f_{X, x-1} - \\
        &\{S_{X, x}(T_{\mathrm{e}}) + \alpha_{X, x}(T_{\mathrm{e}})\}f_{X, x} + \alpha_{X, x+1}(T_{\mathrm{e}})f_{X, x+1}],
    \end{split}
\end{gather}
\noindent
where $f_{X, x}$ is the fraction of the $x$-th ionized atom, $S_{X, x}(T_{\mathrm{e}})$ is the coefficient for the ionization from state $x$ to state $x+1$, and $\alpha_{X, x}(T_{\mathrm{e}})$ is the coefficient for the recombination from state $x+1$ to state $x$ 
\citep[e.g.,][]{hamilton1983}.
We note that the multiple-ionization process, inner shell ionization and subsequent Auger decay, are not included in our calculations.
The rate coefficients are taken from the latest version of AtomDB v3.0.9 \citep{smith2001, foster2012}, which contain the contributions of direct collisional ionization, excitation autoionization, radiative recombination, and dielectronic recombination.

When solving the equations above, we should take into account the time evolution of the electron density caused by the ionization and recombination of the heavy elements as follows:
\noindent
\begin{gather}
    \frac{\mathrm{d} n_{\mathrm{e}}}{\mathrm{d}t} = \sum_X \frac{\mathrm{d} \bar{Z}_X}{\mathrm{d}t} n_X = \sum_X \sum_{x=0}^{X} x \frac{\mathrm{d}f_{X, x}}{\mathrm{d}t} n_{X}.
\end{gather}
\noindent
The ionization process also contributes to the cooling of electrons. 
The kinetic energy of free electrons is consumed to release bound electrons from heavy element ions. 
The electron temperature decrease due to this process can be described as
\noindent
\begin{gather}
    \label{eq: cooling1}
    \frac{\mathrm{d}{\epsilon_{\mathrm{e}}}}{\mathrm{d}t} = - \sum_{X} \sum_{x=0}^{X} I_{X, x} S_{X, x} f_{X, x} n_{\mathrm{X}} n_{\mathrm{e}},
\end{gather}
\noindent
where $I_{X, z}$ is the ionization potential of the $x$-th ionized atom of the atomic number $X$.
We assume that the outermost-shell electron is always released in the collisional ionization process (i.e., the smallest value of $I_{X, z}$ is used).

The coupled equations are numerically integrated by the \texttt{solve\_ivp} function in the Python package Scipy \citep{virtanen2020} with the implicit Runge-Kutta method \texttt{Radau} \citep{hairer1996a} option.
We define an integral valuable $\tau$ (or ``ionization timescale'') as
\noindent
\begin{gather}
    \label{eq: tau}
    \tau = \int_{0}^{t} n_{\mathrm{e}}(t^{\prime})\,\mathrm{d}t^{\prime},
\end{gather}
\noindent
where $t$ is the elapsed time after the shock heating.
To summarize, our model calculates the electron and ion temperatures and ion fractions of each atom simultaneously, as a function of $\tau$ for a given shock velocity and $\beta$ value.

X-ray spectra are synthesized using the atomic database and code provided in PyAtomDB \citep{foster2020}, the Python wrapper of AtomDB \citep{smith2001, foster2012}.
AtomDB calculates the thermal X-ray emission, including continua and emission lines, from optically thin collisional plasmas.
An output of the calcurations is the emissivity $\epsilon$ in a unit of $\mathrm{photon\,cm^{3}\,s^{-1}}$, which can be related to the photon flux $F$ as follows
\begin{align}
    \label{eq: line flux}
    F = \frac{\epsilon\,(T_{\mathrm{e}})}{4 \pi d^{2}} \int n_{\mathrm{e}} n_{\mathrm{H}} \mathrm{d}V,
\end{align}
\noindent
where $\int n_{\mathrm{e}} n_{\mathrm{H}} \mathrm{d}V$ is the volume emission measure and $d$ is the distance to the source.
For the emission line of $i \rightarrow j$ transition of $z$-th ionized element $Z$, the emissivity is calculated as
\begin{align}
    \label{eq: emissivity}
    \epsilon_{Z, z, i \rightarrow j} = A_{Z, z, i \rightarrow j} \frac{n_{Z, z, i}}{n_{\mathrm{e}} n_{Z, z}} \frac{n_{Z, z}}{n_{Z}} \frac{n_{Z}}{n_{\mathrm{H}}}
\end{align}
\noindent
where $n_{Z, z, i}$ is the number density of level $i$ and $A_{Z, z, i \rightarrow j}$ is the Einstein coefficient of the transition.
The level population $n_{Z, z, i}/n_{Z, z}$ is solely determined by the electron temperature and is proportional to $n_{\mathrm{e}}$ under the assumption of coronal approximation.
Therefore, the dependency on electron density vanishes in Equation~\ref{eq: emissivity} \citep[for more details, see][]{phillips2008}, and the X-ray spectrum of our model is determined by $V_{\mathrm{sh}}$, $\beta$, $\tau$, and elemental abundances.

\section{Results} \label{sec: results}

In this section, we calculate the evolution of shock-heated plasmas using the model described in Section~\ref{sec: method}.
We consider two types of plasmas with different compositions: hydrogen-dominant plasmas with solar elemental abundance (\S3.1) and pure-metal plasmas consisting only of Fe ions and electrons (\S3.2). 
The former represents the swept-up medium heated by the SNR forward shock, whereas the latter represents the metal-rich supernova ejecta heated by the reverse shock.

The evolution of the electron and ion temperatures and ion fraction are calculated for each case and compared with predictions of the traditional \texttt{nei} model in XSPEC.
The X-ray spectra are also synthesized and compared with the \texttt{nei} model.

\subsection{Solar abundance case}
\label{subsec: HydrogenRichCase}

\begin{figure}[t]
    \centering
    \includegraphics[width=8.5cm]{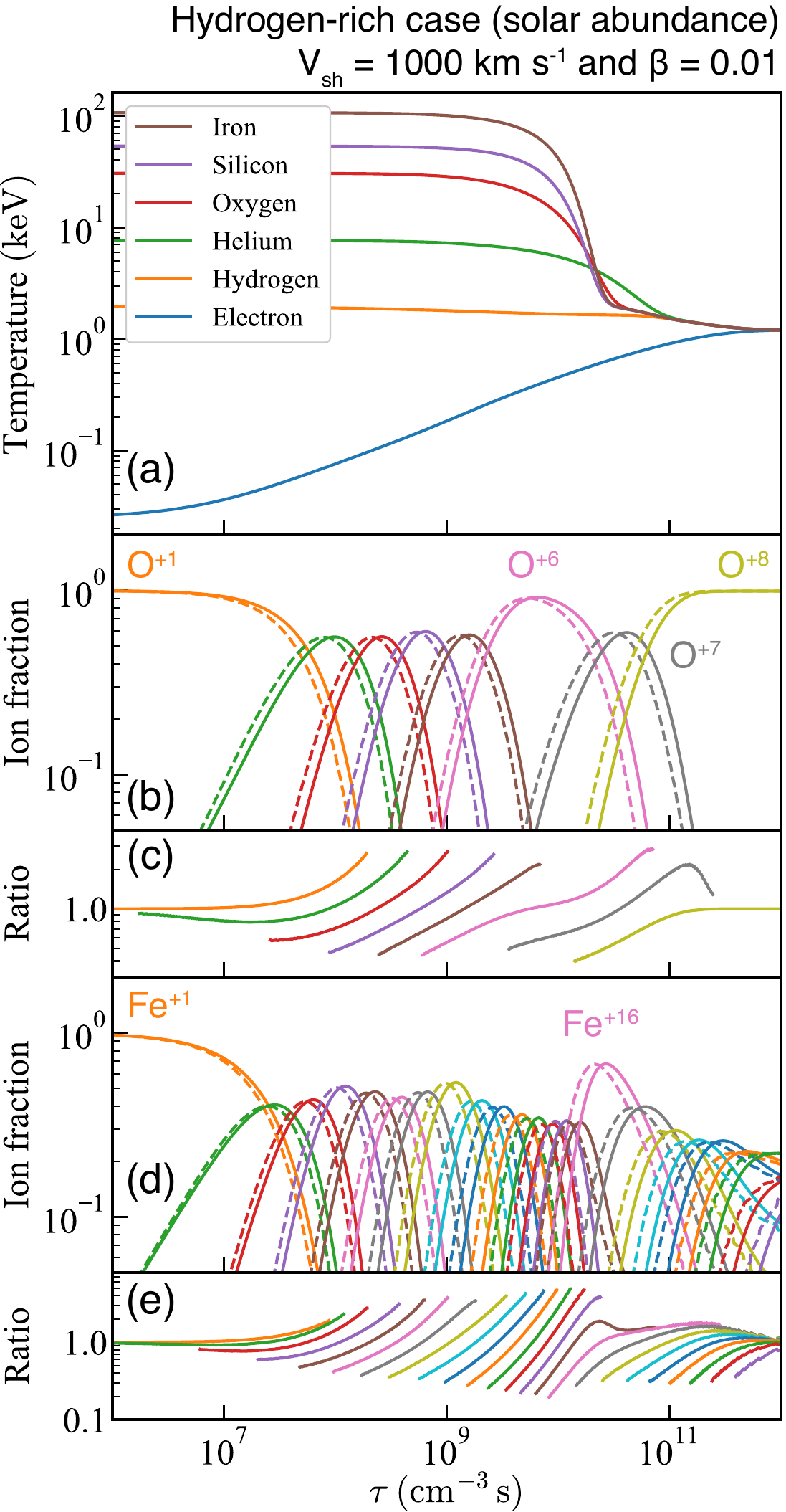}
    \caption
    {
        Temperature and ionization equilibration processes in a solar abundance plasma shock-heated with $V_{\rm sh} =  1000\,\mathrm{km\,s^{-1}}$ and $\beta = 0.01$: 
        (a) Electron and ion temperatures as a function of $\tau$.
        Only the temperatures of several representative elements are plotted, 
        although all the atoms from H to Ni are included in the calculation.
        (b) Ion fraction of oxygen for our model (solid lines) and \texttt{nei} model (dashed lines) and (c) thair ratio as a function of $\tau$.
        The ratios are plotted only when the ion fractions are greater than $10^{-2}$.
        (d) and (e) are the same as (b) and (c), respectively, but for iron.
    }
    \label{fig: HydrogenRichCase}
\end{figure}

We first consider the solar abundance case.
The elemental composition is assumed to be the solar abundances of \cite{Anders_Grevesse1989}.
Figure \ref{fig: HydrogenRichCase}(a) shows the temporal evolution of the ion and electron temperatures, where the shock velocity of $1000\,\mathrm{km\,s^{-1}}$ and $\beta = 0.01$ are assumed as the initial conditions.
The electron temperature gradually increases as $\tau$ increases, and reaches equilibrium with protons at $\tau \sim 10^{12}$\,cm$^{-3}$\,s.
On the other hand, the ion temperatures remain unchanged from the postshock temperatures in the beginning but rapidly decrease at $\tau \sim 10^{10}$\,cm$^{-3}$\,s.
This is because the Coulomb interactions between the ions and protons take place more efficiently when the charge number of the ions becomes higher. 
It is also expected that the higher mass elements reach the equilibrium with protons with a shorter timescale, due to the dependence of $t_{\mathrm{eq, }ij} \propto m_i Z_i^{-2}$ (see Equation \ref{eq: t_eq}).

\begin{figure}[t]
    \centering
    \includegraphics[width=8.5cm]{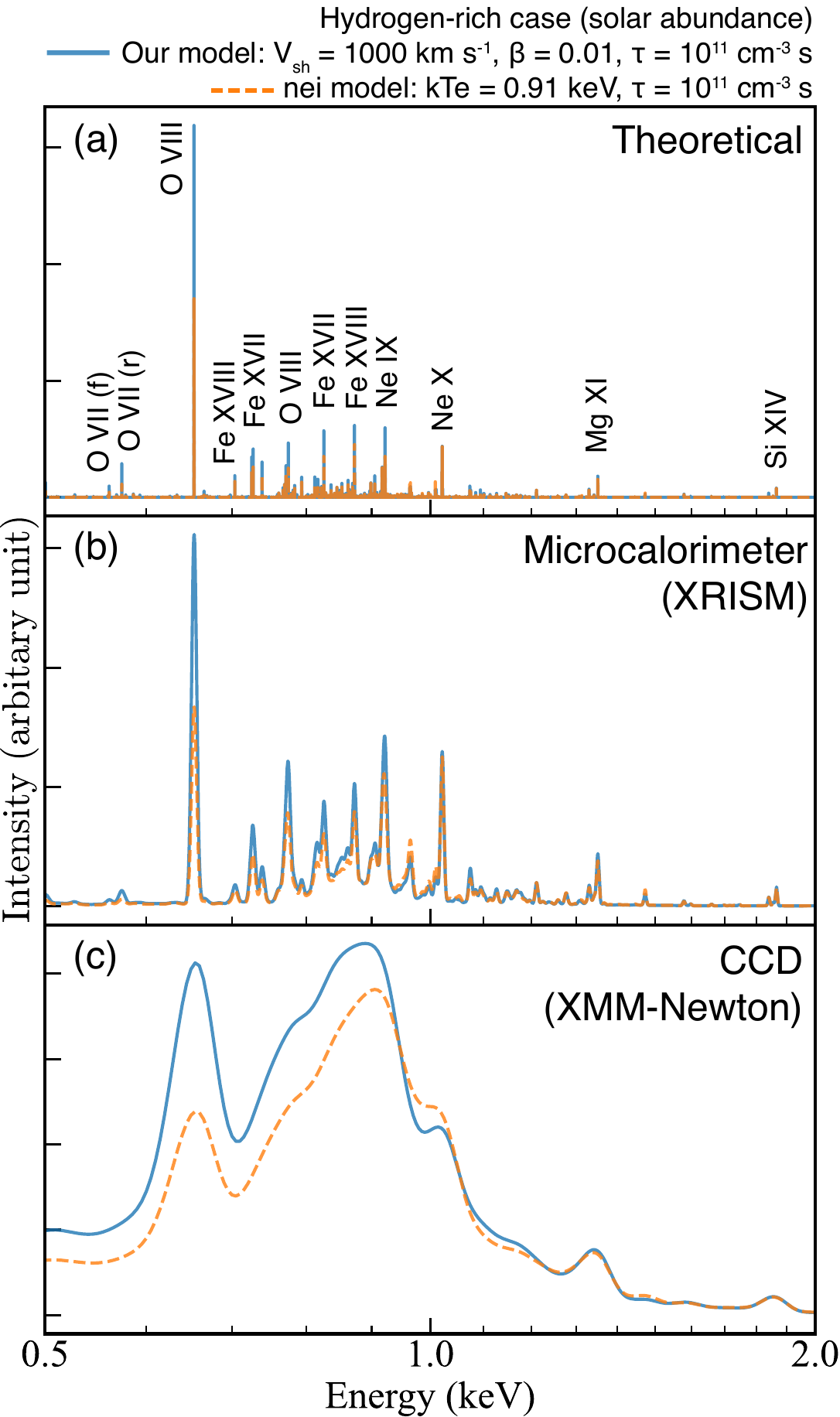}
    \caption
    {
        Synthesized spectra of a solar-abundance plasma in the 0.5--2.0\,keV band, predicted by our model (solid line) and \texttt{nei} (dashed line) model with the identical electron temperature of $0.91\,\mathrm{keV}$ and $\tau$ of $10^{11}\,\mathrm{cm^{-3}\,s}$:
        Panel (a) shows the theoretical spectra, whereas (b) and (c) show the spectra convolved with the instrumental response matrices of XRISM Resolve (Microcalorimeter) and XMM--Newton MOS (CCD), respectively.
    }
    \label{fig: Spectra_HydrogenRichCase}
\end{figure}

Figure~\ref{fig: HydrogenRichCase}(b) shows ion fractions of oxygen as a function of $\tau$,
where our calculations (solid curves) and the AtomDB-based \texttt{nei} model (dashed curves) are compared. 
While our model calculates the evolution of electron temperature and ionization self-consistently, the \texttt{nei} model calculates the ion fractions only using a constant electron temperature and a given $\tau$ value. Therefore, to derive the ion fractions of the \texttt{nei} model at each $\tau$ value, we use the electron temperature corresponding to each $\tau$ in our calcluation shown in Figure~\ref{fig: HydrogenRichCase}(a) (e.g., $kT_{\mathrm{e}} = 0.45\,\mathrm{keV}$ at $\tau = 10^{10}\,\mathrm{cm^{-3}\,s}$ and $kT_{\mathrm{e}} = 0.91\,\mathrm{keV}$ at $\tau = 10^{11}\,\mathrm{cm^{-3}\,s}$). 
Although the ionization degree gradually increases with $\tau$ in both models, our model predicts slower ionization due to the lower electron temperature at the lower $\tau$ regime. 
The ratios of the ion fractions between the two models are plotted in Figure~\ref{fig: HydrogenRichCase}(c), indicating that the fraction of each ion may differ by a factor of up to 3. 
We also show similar plots for iron in 
Figures~\ref{fig: HydrogenRichCase}(d) and (e).
The difference between the two models is even more prominent, especially for the ions holding M-shell electrons.

Figure~\ref{fig: Spectra_HydrogenRichCase}(a) shows synthesized 
X-ray spectra in the 0.5--2.0\,keV band for both models at the identical ionization timescale of 
$\tau = 10^{11}\,\mathrm{cm^{-3}\,s}$. 
Since the initial condition assumed in our model 
(i.e., $V_{\rm sh} = 1000\,\mathrm{km\,s^{-1}}$ and $\beta = 0.01$) corresponds to the electron temperature of 0.91\,keV at $\tau = 10^{11}\,\mathrm{cm^{-3}\,s}$, 
we use this value to synthesize the spectrum based on the \texttt{nei} model. 
In Figures~\ref{fig: Spectra_HydrogenRichCase}(b) and \ref{fig: Spectra_HydrogenRichCase}(c), we show the spectra with the same parameters convolved with the instrumental responses of the Resolve (X-ray microcalorimeter) aboard XRISM and the EPIC/MOS (X-ray CCD) aboard XMM--Newton, respectively.
The difference between the two models is clearly seen even at the energy resolution of the CCD, because of the different ion fractions predicted by the models. For instance, the fraction of O$^{7+}$ ions is predicted to be  0.22\% in our model and 0.11\% in the \texttt{nei} model, resulting in the twice higher O VIII Ly$\alpha$ intensity in the spectrum of the former. 

\subsection{Pure-iron case}
\label{subsec: PureIronCase}

\begin{figure}[t]
    \centering
    \includegraphics[width=8.5cm]{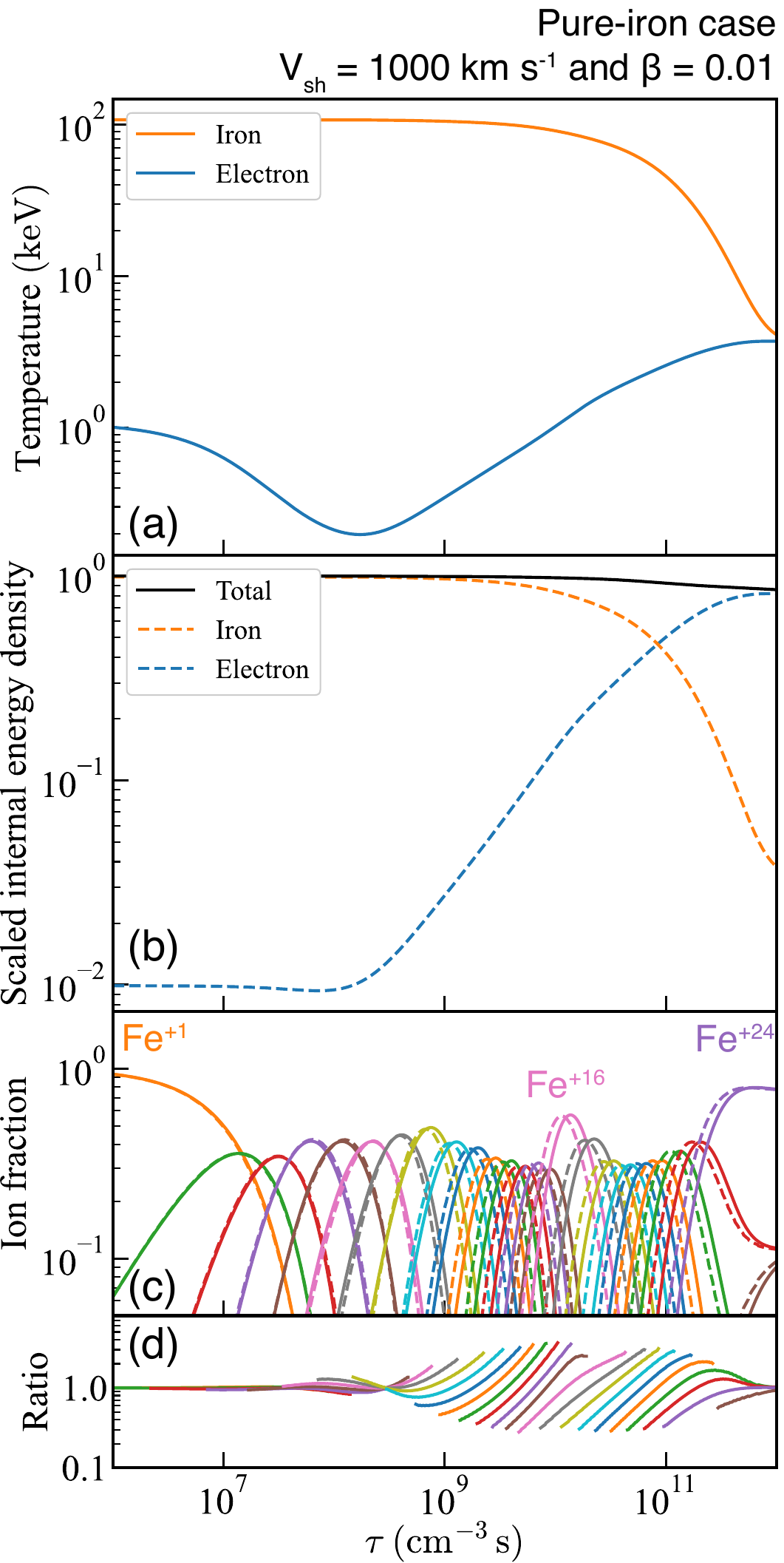}
    \caption
    {
        Temperature and ionization equilibration processes in a pure-iron plasma shock-heated with $V_{\rm sh} =  1000\,\mathrm{km\,s^{-1}}$ and $\beta = 0.01$: 
        (a) Electron and iron temperatures as a function of $\tau$.
        (b) Total internal energy density (solid lines) and contribution from each species (dased lines) as a function of $\tau$, with the total scaled to unity at $\tau = 0$.
        (c) Ion fraction of iron for our model (solid lines) and \texttt{nei} model (dashed lines) and (d) their ratio as a function of $\tau$.
        The ratios are plotted only when the ion fractions are greater than $10^{-2}$.
    }
    \label{fig: PureIronCase}
\end{figure}

\begin{figure}[t]
    \centering
    \includegraphics[width=8.5cm]{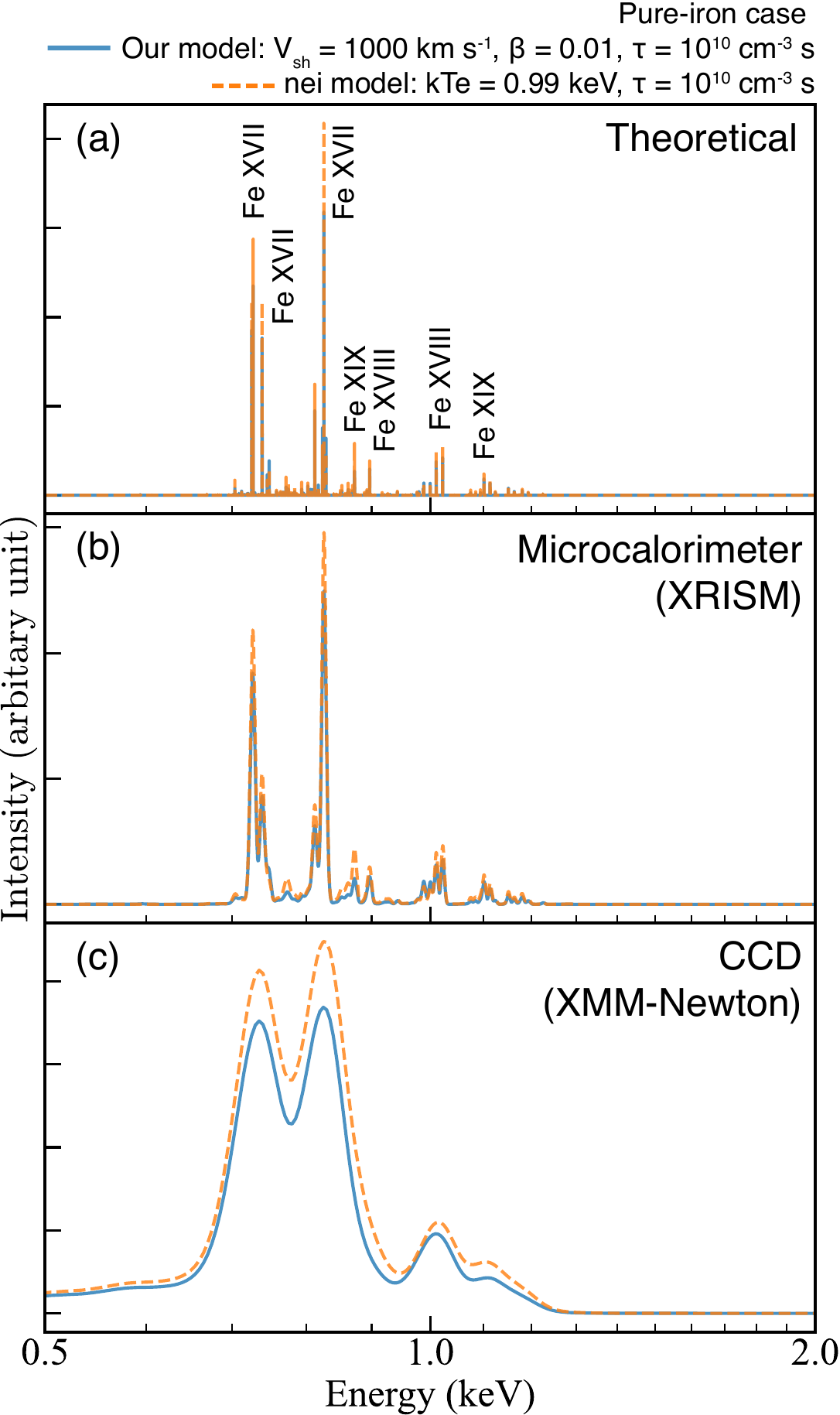}
    \caption
    {
        Same as Figure 3 but for a pure-iron plasma.
    }
    \label{fig: Spectra_pureiron}
\end{figure}

SNRs contain heavy elements generated during the stellar evolution and supernova explosion.
Therefore, the ejecta heated by the SNR reverse shock may consist of heavy elements and electrons with no mixture of hydrogen or helium. 
In such ``metal-rich'' plasma, free electrons are provided by heavy elements, 
and thus the temporal evolution of the plasma properties is different from that in the hydrogen-dominant plasma. 
Here we consider a plasma composed purely of iron as the simplest example of the metal-rich case.

We calculate the temporal evolution of a pure-iron plasma heated by a shock wave with a velocity of $1000\,\mathrm{km\,s^{-1}}$ and the collisionless electron heating of $\beta = 0.01$.
Figure~\ref{fig: PureIronCase}(a) shows the electron and iron temperatures as a function of $\tau$.
In contrast to the hydrogen-dominant case, the electron temperature does not monotonically increase but decreases at the lower $\tau$ regime.
As shown in Figure~\ref{fig: PureIronCase}(b), the internal energy density of the electrons is constant at that regime. Therefore, the decrease in electron temperature is caused by the increase in the electron number density due to ionization.

Figure~\ref{fig: PureIronCase}(b) shows the ion fractions of Fe predicted by our model and the \texttt{nei} model.
Again, the latter assumes a constant electron temperature to calculate the ion fractions.
The ratio between the two models is shown in Figure~\ref{fig: PureIronCase}(c). 
At the lower $\tau$ regime, the ratio shows a different trend from the hydrogen-dominant case due to the decrease in electron temperature in our model.
The difference between the models is significant in the range $\tau = 10^{8} \text{--} 10^{12}\,\mathrm{cm^{-3}\,s}$, by a factor of up to 4. 

The spectra synthesized for both models at $\tau = 10^{10}\,\mathrm{cm^{-3}\,s}$ are calculated and shown in Figure~\ref{fig: Spectra_pureiron}.
In the pure-iron case, the initial condition of 
$V_{\rm sh} = 1000\,\mathrm{km\,s^{-1}}$ and $\beta = 0.01$
lead to the electron temperature of $0.99\,\mathrm{keV}$ at $\tau = \red{10^{10}}\,\mathrm{cm^{-3}\,s}$. 
Therefore, we use this temperature to generate the synthesized spectrum of the \texttt{nei} model. 
The major difference is seen in the energies of the Fe XVII lines, due to the difference in the fraction of Fe$^{16+}$ ions between the models.

\section{Discussion} \label{sec: discussion}

\subsection{Systematic bias in parameter estimation using the \texttt{nei} model}
\label{subsec: systematic bias}

\begin{figure*}[t]
    \centering
    \includegraphics[width=18.0cm]{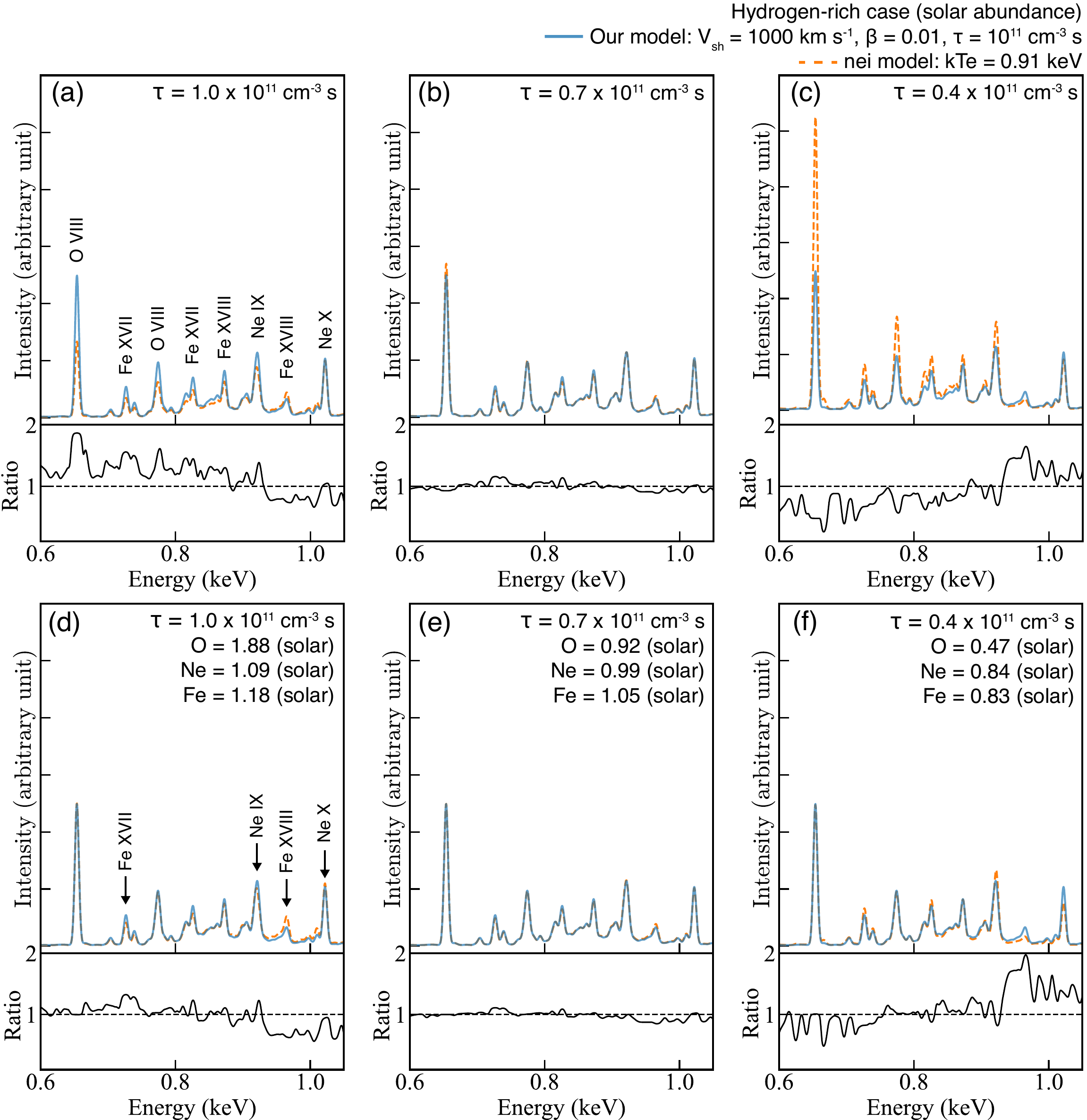}
    \caption
    {
        Synthesized \texttt{nei} spectra of a solar-abundance plasma in the 0.6--1.05~keV band for varying $\tau$ and abundances.
        The blue solid lines show the synthesized spectrum predicted by our model with $V_{\mathrm{sh}} = 1000\,\mathrm{km\,s^{-1}}$, $\beta = 0.01$, and $\tau = 10^{11}\,\mathrm{cm^{-3}\,s}$.
        The orange dashed lines show the synthesized spectrum predicted by our model with an electron temperature of $0.91\,\mathrm{keV}$ and varying $\tau$ values.
        The $\tau$ values for the \texttt{nei} model are $1.0 \times 10^{11}\,\mathrm{cm^{-3}\,s}$ for (a) and (d), $0.7 \times 10^{11}\,\mathrm{cm^{-3}\,s}$ for (b) and (e) and $0.4 \times 10^{11}\,\mathrm{cm^{-3}\,s}$ for (c) and (f).
        The abundances for the \texttt{nei} model are solar abundances for (a), (b), and (c), and optimized by the least squares method for (d), (e), and (f).
        The ratios between the two models are shown at the bottom of each panel.
        Each spectrum is convolved with the instrumental response matrices of XRISM Resolve.
    }
    \label{fig: BiasCorrectedSpectra}
\end{figure*}

In the previous section, we calculated the temporal evolution of the electron/ion temperatures and ion fractions self-consistently. 
Comparing the results with the NEI calculations of the traditional constant-temperature models, we have found that the traditional \texttt{nei} model predicts higher ionization than our model at a certain ionization timescale ($\tau$), implying that spectral modeling using the \texttt{nei} model may lead to a systematic bias in the estimate of the plasma age. 

Here we assess this bias more quantitatively. 
Figure~\ref{fig: BiasCorrectedSpectra}(a), (b), and (c) show synthesized spectra in the 0.6--1.05\,keV band for the \texttt{nei} model with an electron temperature of 0.91\,keV and different $\tau$ values ($1.0 \times 10^{11}$, $0.7 \times 10^{11}$, and $0.4 \times 10^{11}$\,cm$^{-3}$\,s).
The spectrum predicted by our model for $V_\mathrm{sh} = 1000\,\mathrm{km\,s^{-1}}$, $\beta = 0.01$, and $\tau = 1.0 \times 10^{11}\,\mathrm{cm^{-3}\,s}$ is also shown in each panel for reference.
We find that the \texttt{nei} model spectrum best reproduces the reference spectrum at $\tau = 0.7 \times 10^{11}\,\mathrm{cm^{-3}\,s}$, indicating that the \texttt{nei} model underestimates $\tau$ by $\sim$\,30\%.

Although the \texttt{nei} model with a lower $\tau$ value moderately reproduces the reference spectrum, a marginal discrepancy remains between the two models, which leads to another systematic bias in the abundance estimate.
As the derivation of elemental abundances is one of the major goals of X-ray observations of SNRs, it is also important to assess the systematic bias in their estimation.

In Figure~\ref{fig: BiasCorrectedSpectra}(d), (e), and (f), we show the spectra with the same $\tau$ but with various abundances of O, Ne, and Fe.
The abundances are fitted using the least squares method to best match our model spectrum.
In Figure~\ref{fig: BiasCorrectedSpectra}(d) and (f), despite significant adjustments to the abundances, the intensity ratios between ions with different charge states (e.g., Fe XVII/Fe XVIII and Ne IX/Ne X) in the reference spectrum do not align with those in the \texttt{nei} spectrum due to differing ion fraction ratios.
On the other hand, in Figure~\ref{fig: BiasCorrectedSpectra}(e), the reference spectrum can be well reproduced after the abundance corrections of less than 10\%.
We note that the adjustment of the abundances does not improve the shape of the residuals between the two models. Therefore, to reduce the discrepancy between the models, $\tau$ always needs to be corrected.

In Section~\ref{subsec: PureIronCase}, the pure-iron plasma was considered as the simplest case of ejecta-dominant plasma.
As shown in Figure~\ref{fig: PureIronCase}(a), the electron temperature decreases at lower $\tau \sim 10^{8}\,\mathrm{cm^{-3}\,s}$ due to efficient ionization cooling in the pure-iron plasma.
The drop in the electron temperature changes the emissivity of the emission lines and the ion fractions, which in turn affects the line fluxes (Equation~\ref{eq: line flux}).
Therefore, a proper treatment of the ionization cooling effect is crucial for interpreting the emission lines from the ejecta-dominant plasma with low ionization states \citep[e.g, for optical study, see][]{seitenzahl2019}.

\subsection{Dependency of the temperatures on shock velocity and $\beta$}
\label{subsec: parameter dependency}

\begin{figure*}[t]
    \centering
    \includegraphics[width=18.0cm]{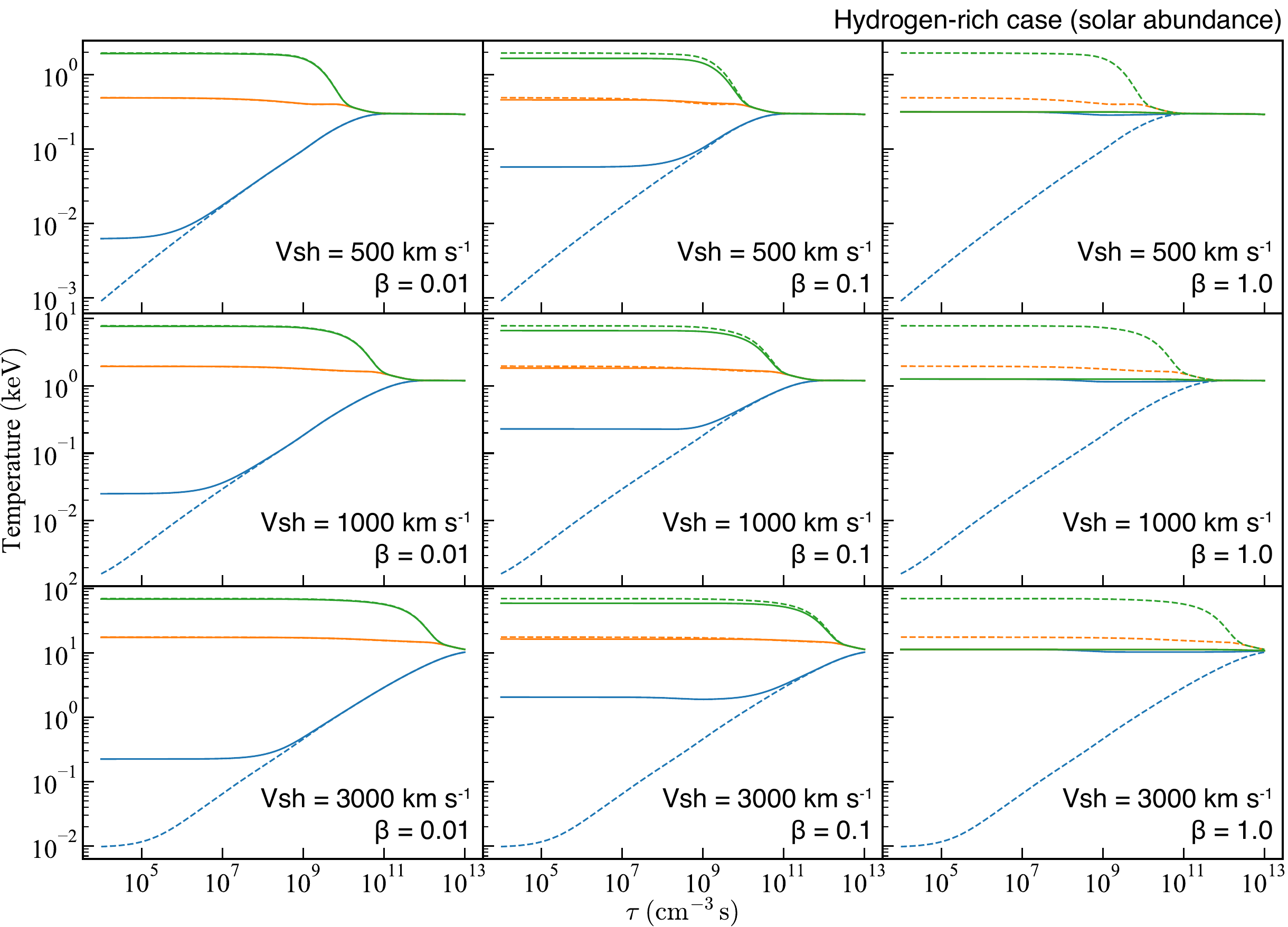}
    \caption
    {
        Temperature equilibration processes in a solar abundance plasma for varying initial conditions.
        The simulations used $V_{\mathrm{sh}} = 500\,\mathrm{km\,s^{-1}}$, $1000\,\mathrm{km\,s^{-1}}$, and $3000\,\mathrm{km\,s^{-1}}$, and $\beta = $ 0.01, 0.1, and 1.0, resulting in a 3 $\times$ 3 array of subplots.
        The dashed line in each subplot shows the case with the same shock velocity but $\beta = m_{\mathrm{e}}/\mu m_{\mathrm{p}}$ for reference.
    }
    \label{fig: CollisionlessHeating}
\end{figure*}

In this subsection, we investigate how the temperature evolution in the shock-heated plasma is affected by the initial conditions, i.e., $V_{\mathrm{sh}}$ and $\beta$.
Figure \ref{fig: CollisionlessHeating} indicates the electron and ion temperatures as a function of $\tau$ for solar-abundance plasmas with different shock velocities (500, 1000, and $3000\,\mathrm{km\,s^{-1}}$) and initial electron-to-ion temperature ratios (0.01, 0.1, and 1.0).
The dashed lines in each panel indicate the model with the minimal beta (i.e., $\beta = m_{\mathrm{e}}/\mu m_{\mathrm{p}}$) for reference.
We find that the timescale of temperature equilibrium depends on the shock velocity, as predicted by Equations~\ref{eq: shock heating} and \ref{eq: t_eq}.
When the shock velocity exceeds $1000\,\mathrm{km\,s^{-1}}$, the equilibrium timescale exceeds $10^{12}\,\mathrm{cm^{-3}\,s}$, a typical value of ionization equilibration timescale \citep{masai1984, smith2010}.
Therefore, the systematic biases discussed in the previous subsection are more important for a faster shock.
It is worth noting that, although different $\beta$ values lead to different initial electron temperatures, their trajectories merge before the plasma reaches the thermal equilibrium, and thus the timescale for the thermal equilibrium does not depend on $\beta$.
Because of this, it is difficult to constrain the collisionless electron heating efficiency from an X-ray spectrum of NEI plasma with a relatively high $\tau$ value.

\subsection{Thermal broadening of X-ray emission lines}

\begin{figure}[t]
    \centering
    \includegraphics[width=8.5cm]{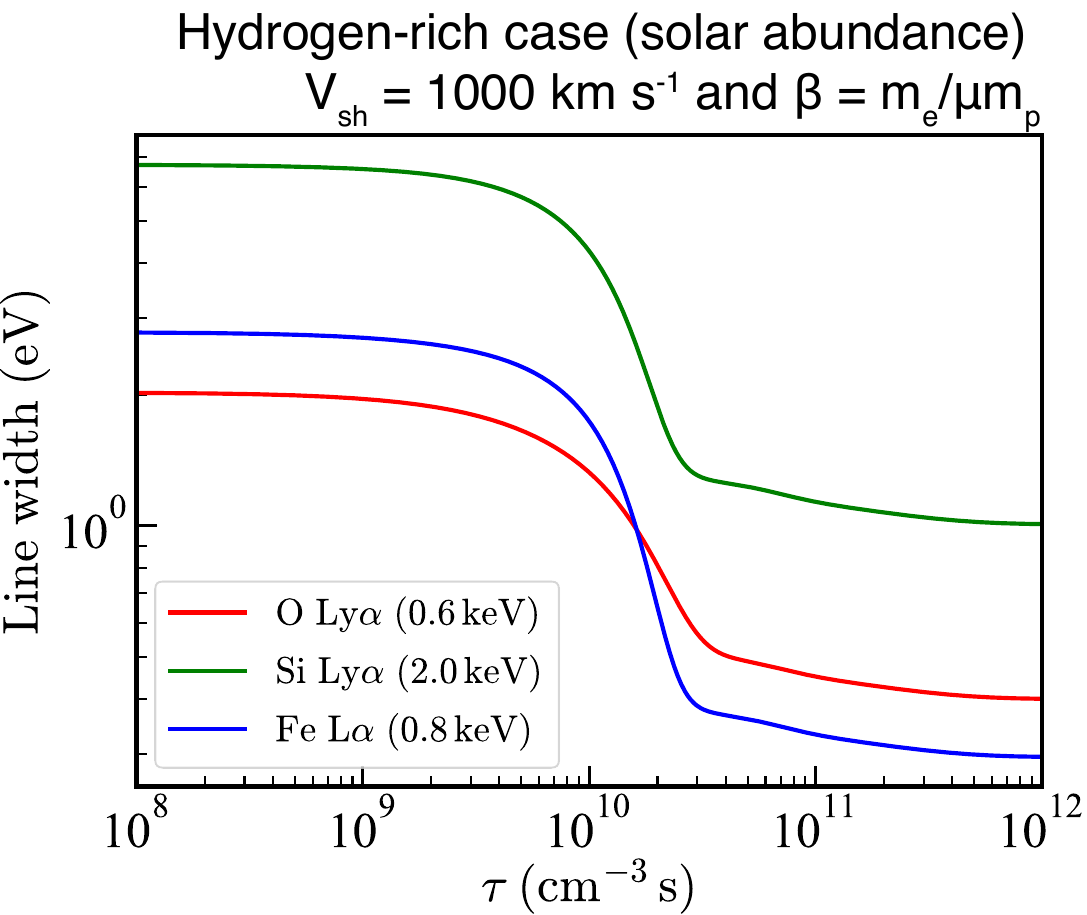}
    \caption
    {
        Thermal broadening width in FWHM of oxygen Ly$\alpha$ (Red), silicon Ly$\alpha$ (Green), and iron L$\alpha$ (blue) in the solar-abundance plasma shock heated with $V_{\rm sh} = 1000\,\mathrm{km\,s^{-1}}$ and $\beta = m_{\mathrm{e}}/\mu m_{\mathrm{p}}$ as a function of $\tau$.
    }
    \label{fig: ThermalBroadening}
\end{figure}

As mentioned in Section~\ref{sec: intro}, high-resolution X-ray spectroscopy provides us with unique constraints on shock-heated plasmas.
One such example is a direct measurement of ion temperature, the key to estimating shock velocity, which in turn unravels the shock heating mechanism.
Since shock-heated ions maintain high temperatures until their equilibrium (see Figure~\ref{fig: HydrogenRichCase}), emission lines from these ions are broadened due to their thermal motion.
The relationship between the full width at half maximum (FWHM) of the broadened line $\Delta (h \nu)$ and ion temperature $kT_i$ is given by
\noindent
\begin{gather}
    \label{eq: thermal broadening}
    \frac{\Delta (h \nu)}{h \nu_0} = 2\sqrt{\ln{2}}\left(\frac{2kT_i}{m_i c^2}\right)^{1/2},
\end{gather}
\noindent
where $h \nu_0$ is the rest frame energy of the emission line.

We consider the solar-abundance case presented in  Section~\ref{subsec: HydrogenRichCase} (i.e., $V_{\mathrm{sh}} = 1000\,\mathrm{km\,s^{-1}}$, $\beta = 0.01$, and $\tau = 10^{11}\,\mathrm{cm^{-3}\,s}$), and convert ion temperatures into line width using Equation~\ref{eq: thermal broadening}.
The FWHM of O K-shell lines ($h \nu_0 = 0.6\,\mathrm{keV}$), Si K-shell lines ($h \nu_0 = 2.0\,\mathrm{keV}$), and Fe L-shell lines ($h \nu_0 = 0.8\,\mathrm{keV}$) are shown in Figure~\ref{fig: ThermalBroadening}.
As the FWHM holds its initial value at a low $\tau \lesssim 10^{9}\,\mathrm{cm^{-3}\,s}$, the line widths of optical and UV emission, which are thought to originate from immidiate post shock regions, can be used to constrain the initial ion temperautre and thus the shock velocity.
On the other hand, the FWHM decreases rapidly when the temperature equilibrium is achieved ($\tau \gtrsim 10^{10}\,\mathrm{cm^{-3}\,s}$), and the line width is smaller by an order of magnitude than that of the initial width.
Therefore, the initial conditions including the shock velocity can no longer be constrained from the line width, once the thermal equilibrium is achieved.

Such a situation is likely observed in SN 1987A, where the forward shock is propagating into a dense environment \citep{sugerman2005}, and thus thermal equilibrium is partially achieved despite its young age.
Recently, \cite{miceli2019} measured the width of X-ray emission lines and compared them with their three-dimensional hydrodynamical simulations.
The observed line width was broader than expected from the hydrodynamical model calculation that took into account the kinematic Doppler broadening.
They attributed the excess in width to the thermal motion of ions, and concluded that the ion temperatures are mass-proportional. 
In their hydrodynamical model, however, a high $\tau$ value ($\gtrsim 3 \times 10^{11}\,\mathrm{cm^{-3}\,s}$) is expected since the forward shock propagates into the dense equatorial ring ($\sim 10^{3}\,\mathrm{cm^{-3}}$) and clumps ($\sim 10^{5}\,\mathrm{cm^{-3}}$) and the elapsed time since ring heating ($\sim 10\,\mathrm{yr}$).
In particular, the plasma component originating from the dense material is expected to have an electron temperature of $\sim 0.5\,\mathrm{keV}$ and the highest emission measure than any other components (see Figure 7 of \citeauthor{orlando2015}~\citeyear{orlando2015}, which is the basis of the hydrodynamic model used in \citeauthor{miceli2019}~\citeyear{miceli2019}), implying that the observed emission lines such as Fe XVII are dominated by this component.
A high $\tau$ value was also suggested by several observations \citep[e.g.,][]{zhekov2009, bray2020, ravi2021}.
Our results show the timescale of temperature equilibrium is one order of magnitude less than these values, $\tau \sim 2 \times 10^{10}\,\mathrm{cm^{-3}\,s}$ (see Figure~\ref{fig: ThermalBroadening}), for the shock velocity of $1000\,\mathrm{km\,s^{-1}}$.
This fact suggests that at least part of the X-ray emitting plasma in SN 1987A has already reached thermal equilibrium, implying a smaller contribution of thermal broadening to the observed line width.
Therefore, the observed broad line width could be explained by an improvement of the hydrodynamical model (kinematic broadening) rather than the presence of thermal broadening.
In order to properly interpret the broad line width of SN 1987A, it is necessary to consider each contribution of thermal broadening and kinematic broadening.
At this point, \cite{sapienza2024} simulates the X-ray spectrum of SN 1987A with the XRISM observation to be conducted in 2024.
They predict that at the time of the XRISM observations, the ejecta emission heated by the reverse shock with a velocity of $\sim 3000\,\mathrm{km\,s^{-1}}$ is enhanced and thus kinematic broadening is dominant in the emission lines.
Therefore, the XRISM observation will purely constrain the dynamics of SN 1987A.

\section{Application to Observational Data} \label{sec: Application}

In this section, we apply our model to actual observations to confirm that reasonable physical parameters can be constrained by spectral modeling.
We also compare the results with those obtained with the conventional \texttt{nei} model and examine if the differences are consistent with our expectations.
First, an implementation of our model in XSPEC \citep{Arnaud1996}, the standard software for X-ray spectral analysis, is described in Section~\ref{subsec: implementation}.
We then apply our model to X-ray spectra of N132D, the X-ray brightest SNR in the Large Magellanic Cloud (LMC), obtained by deep Chandra observations with a total effective exposure of $\sim 900$~ks \citep{plucinsky2018}.
We consider this object to be 
one of the most suitable targets to apply our model for the following reasons:
(1) The forward shock structure is clear in the south rim \citep{sharda2020}.
(2) X-ray emission from regions close to the shock front is dominated by the swept-up ISM \citep{Suzuki2020,sharda2020}.
(3) There is little contribution of non-thermal components to the X-ray spectrum of the ISM \citep{bamba2018}.
A combination of the excellent angular resolution ($0\farcs5$ in half-power diameter) and deep exposure of the Chandra observations allows us to extract spectra with high photon statistics from regions just behind the forward shock.
In this paper, we only demonstrate the capability of our model to estimate physical parameters.
The scientific implications of the presented results of this application will be discussed in a separate paper (Okada et al. in prep.).
The uncertainties quoted in the text and table and the error bars in the figures represent a 1$\sigma$ confidence level.

\subsection{Implementation of our model in XSPEC}
\label{subsec: implementation}

Our model can be implemented in XSPEC \citep{Arnaud1996} version 12.13.1 by constructing an ``atable'' format file that stores model spectra pre-computed on a grid of the three-dimensional parameter space defined by 
$V_{\mathrm{sh}}$ (ranging from 500 to $5000\,\mathrm{km,s^{-1}}$),
$\tau$ (from $10^{8}$ to $10^{13}\,\mathrm{cm^{-3},s}$), and 
$\beta$ (from $m_{\mathrm{e}}/m_{\mathrm{p}}$ to 1). 
There are 50 grid points for each parameter.
The grid for shock speed is linearly spaced, while the grids for the other parameters are logarithmically spaced.
At each grid point, the spectrum is synthesized using our model with the corresponding values of $V_{\mathrm{sh}}$, $\beta$, and $\tau$.
The synthesized spectra have 3000 linearly spaced energy bins ranging from 0.1 keV to 15.0 keV, and emissions from individual ions as separate components to allow for variations in abundances.
The synthesized spectrum is then stored in each grid of the parameter space.
We named this model as ``IONization and TEmperature Non-equilibrium Plasma model'' (\texttt{IONTENP} model).
Free parameters of the \texttt{IONTENP} model are $V_{\mathrm{sh}}$, $\tau$, $\beta$, relative abundances to hydrogen, and normalization.
The normalization is defined as $\frac{10^{-14}}{4 \pi D^2} \int n_{\mathrm{e}} n_{\mathrm{H}} \mathrm{d}V$ where $D$ is the distance to the source, $n_{\mathrm{e}}$ is the electron number density, and $n_{\mathrm{H}}$ is the hydrogen number density, which is the same as the \texttt{nei} model in \texttt{XSPEC}.
Given these parameters, the model predicts the X-ray spectrum at the specified value of $\tau$ by interpolating between the grid points.
We note that only the energy transfers among electrons, proton, and helium nuclei are calculated to reduce computational complexity.
This approximation is reasonable enough to calculate the electron temperature since the number density of the heavy elements is much lower than that of the light elements in the typical ISM. 

\subsection{Spectral analysis}

\begin{figure}[t]
    \centering
    \includegraphics[width=8.5cm]{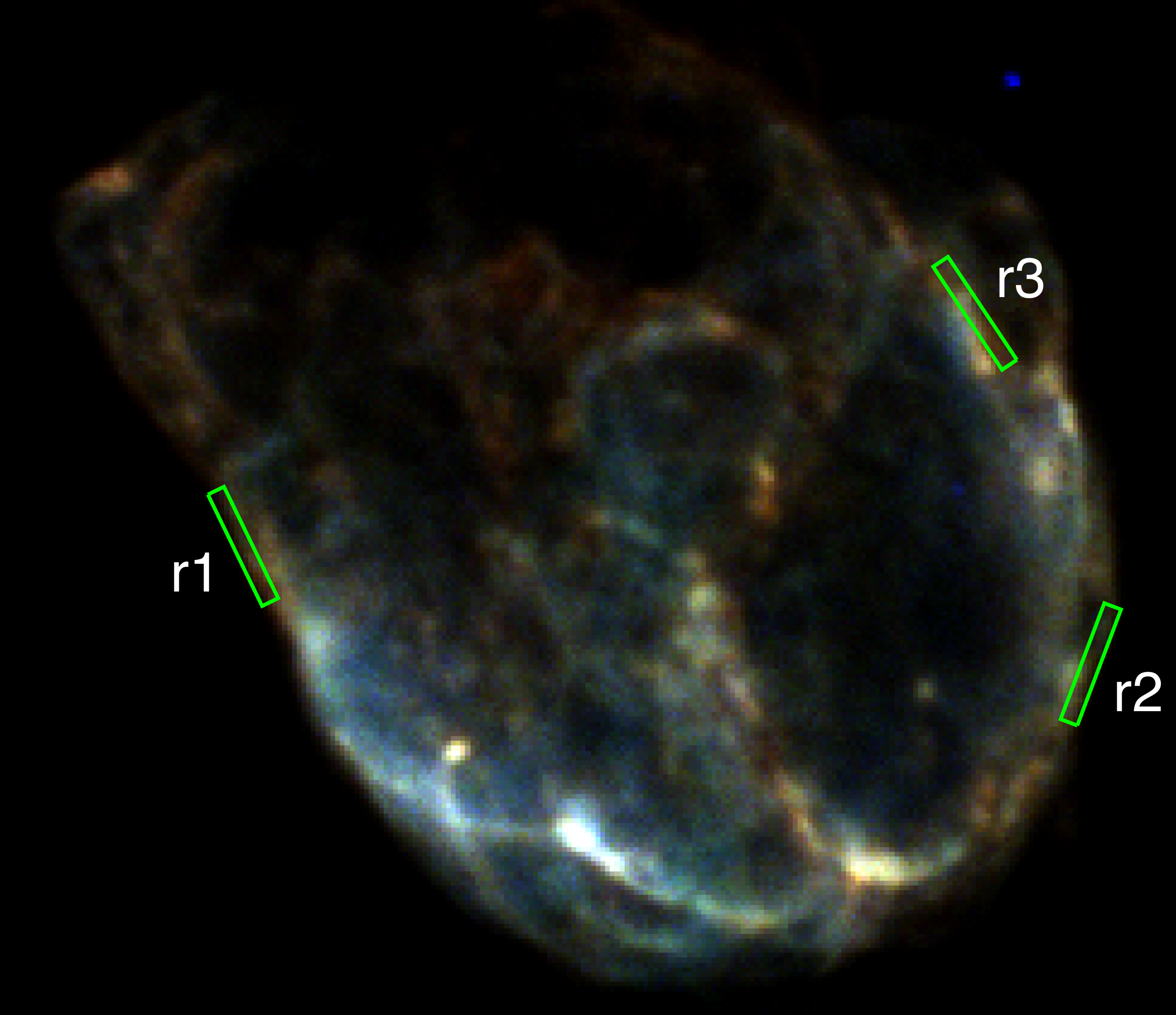}
    \caption
    {
        A three-color ACIS flux image of N132D. Red, blue, and green correspond to 0.5-1.2 keV, 1.2-2.0, and 2.0-7.0 keV, respectively.
    }
    \label{fig: N132D_image}
\end{figure}

\begin{figure}[t]
    \centering
    \includegraphics[width=8.5cm]{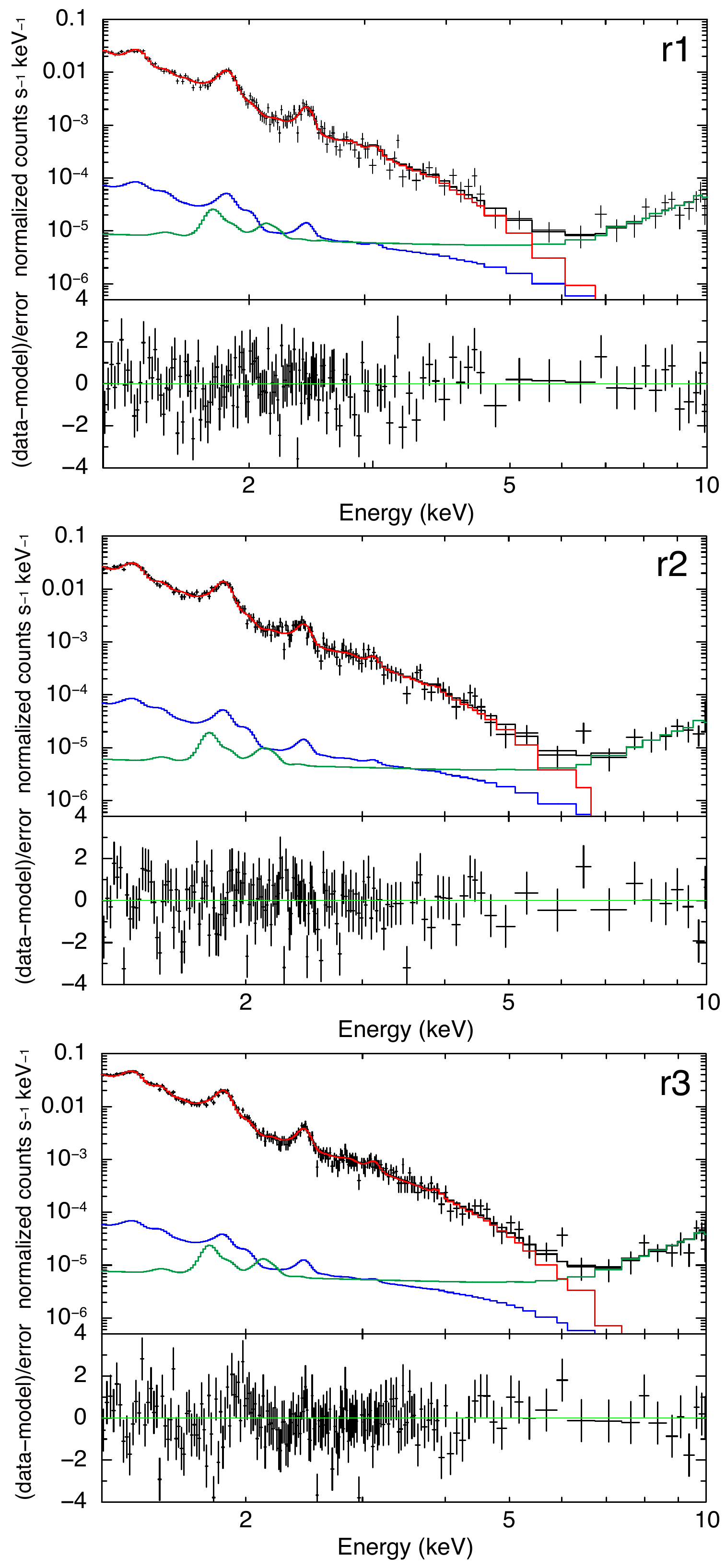}
    \caption
    {
        The Chandra ACIS-S source spectra of N132D and the best-fit models.
        The red, green and blue lines show the thermal emission of the \texttt{IONTENP} model, detector background, and sky background, respectively.
    }
    \label{fig: source_spectrum}
\end{figure}

In this section, we analyze the Chandra data of N132D using the \texttt{IONTENP} model described in the previous section.
We reprocess the data following the standard procedure and obtain a total effective exposure of 867 ks.
The details of the data reprocessing are described in Appendix~\ref{appendix: observation}.

Figure~\ref{fig: N132D_image} shows a three-color, exposure-corrected ACIS image of N132D.
Three source regions are defined as rectangles with a size of $0\farcm21 \times 0\farcm03$ with the major axes parallel to the shock front (see Figure~\ref{fig: N132D_image}).
The background region is a rectangle centered at (RA, Dec) = (81.2430, -69.6619) with a size of $1\farcm67 \times 0\farcm33$ and an angle of 22 degrees measured counterclockwise from the north.
Since the roll angle is different for each observation, we confirm that the background region is selected to fall within the ACIS-S3 chip for all the observations.
The spectral fit is performed using XSPEC version 12.13.1 \citep{Arnaud1996} based on the C-statistic \citep{Cash1979} for unbinned spectra using the solar abundance table of \cite{Anders_Grevesse1989}.
The spectra of all the observations are merged using the CIAO tool \texttt{combine\_spectra} to improve the photon statistics.
We exclude data below 1.2 keV in order to minimize systematic uncertainties due to the time-dependent quantum efficiency by the increasing contamination \citep{plucinsky2018a}.

\begin{table*}[t]
    \caption{Best-fit parameters for the source spectra of N132D}
    \label{Tab: best-fit of source}
    \begin{tabular}{cccccccc}
        \hline \hline
            \texttt{IONTENP} & $V_{\mathrm{sh}}\,\mathrm{(km\,s^{-1})}$ & $\tau\,\mathrm{(10^{11}\,cm^{-3}\,s)}$ & Mg & Si & S & norm\tablenotemark{a} $\mathrm{(10^{-4}\,cm^{-5})}$ & cstat/dof \\ \hline
            r1 & $816^{+20}_{-16}$ & $2.34^{+0.28}_{-0.27}$ & $0.38^{+0.03}_{-0.03}$ & $0.31^{+0.02}_{-0.02}$ & $0.44^{+0.06}_{-0.05}$ & $3.24^{+0.19}_{-0.18}$ & 267.83/287 \\
            r2 & $884^{+22}_{-18}$ & $1.96^{+0.21}_{-0.20}$ & $0.42^{+0.03}_{-0.03}$ & $0.36^{+0.02}_{-0.02}$ & $0.29^{+0.05}_{-0.04}$ & $3.06^{+0.15}_{-0.13}$ & 267.19/278 \\
            r3 & $879^{+13}_{-11}$ & $2.62^{+0.21}_{-0.20}$ & $0.40^{+0.02}_{-0.02}$ & $0.31^{+0.02}_{-0.02}$ & $0.33^{+0.04}_{-0.03}$ & $4.67^{+0.14}_{-0.16}$ & 289.74/311 \\ \hline \hline
            \texttt{nei} & $kT_{\mathrm{e}}$ (keV) & $\tau\,\mathrm{(10^{11}\,cm^{-3}\,s)}$ & Mg & Si & S & norm\tablenotemark{a} $\mathrm{(10^{-4}\,cm^{-5})}$ & cstat/dof \\
            \hline
            r1 & $0.75^{+0.03}_{-0.02}$ & $2.10^{+0.34}_{-0.29}$ & $0.26^{+0.03}_{-0.02}$ & $0.21^{+0.02}_{-0.02}$ & $0.30^{+0.04}_{-0.04}$ & $5.02^{+0.47}_{-0.50}$ & 268.07/287 \\
            r2 & $0.88^{+0.02}_{-0.03}$ & $1.51^{+0.20}_{-0.16}$ & $0.37^{+0.03}_{-0.02}$ & $0.42^{+0.02}_{-0.02}$ & $0.23^{+0.03}_{-0.03}$ & $4.92^{+0.26}_{-0.23}$ & 273.74/278 \\
            r3 & $0.88^{+0.02}_{-0.01}$ & $2.09^{+0.19}_{-0.20}$ & $0.35^{+0.02}_{-0.02}$ & $0.35^{+0.02}_{-0.02}$ & $0.25^{+0.03}_{-0.02}$ & $8.10^{+0.31}_{-0.25}$ & 282.72/311 \\ \hline
        \end{tabular}
    \tablenotetext{a}{Normalization is defined as $\frac{10^{-14}}{4 \pi D^2} \int n_{\mathrm{e}} n_{\mathrm{H}} \mathrm{d}V$, where $D$ is the distance to the source, $n_{\mathrm{e}}$ is the electron number density, and $n_{\mathrm{H}}$ hydrogen number density.}
\end{table*}

The spectral model for the source regions consists of the source and background components. Details of the background modeling are described in Appendix~\ref{appendix: background}.
To fit the source spectra, we introduce two models, \texttt{tbnew*tbnew*(IONTENP)} and \texttt{tbnew*tbnew*(nei)}.
The two \texttt{tbnew}\footnote{\url{https://pulsar.sternwarte.uni-erlangen.de/wilms/research/tbabs/}} components are for the foreground absorption in the Milky Way and the LMC, using solar abundances for the Milky Way component and LMC-mean value \citep{dopita2019} for the LMC component, respectively.
The hydrogen column densities of the Milky Way and LMC components are fixed to $N_{\mathrm{H,\, Gal}} = 5.5 \times 10^{20}\,\mathrm{cm^{-2}}$ \citep{dickey1990} and $N_{\mathrm{H,\, LMC}} = 10.0 \times 10^{20}\,\mathrm{cm^{-2}}$ \citep{sharda2020}.
The free parameters of the \texttt{IONTENP} are the shock speed ($V_{\mathrm{sh}}$), $\tau$, normalization, and abundances of Mg, Si, and S.
We fix $\beta$ at the minimum value, because several previous studies have shown sufficiently high $\tau$ values in N132D \citep{sharda2020}, indicating that the effect of beta is negligible as discussed in Sec~\ref{subsec: parameter dependency}.
Abundances of other elements are fixed to the LMC value \citep{dopita2019} or 0.3 if no value is available in the literature.
The parameters of \texttt{nei} are, on the other hand, the electron temperature ($kT_{\mathrm{e}}$), $\tau$, normalization, and abundances of Mg, Si, and S.
The source spectra are well described by both models, and the best-fit spectra and parameters are given in Figure~\ref{fig: source_spectrum} and Table~\ref{Tab: best-fit of source}, respectively.

\subsection{Discussion on the spectral analysis}

The result of the spectral fitting based on the \texttt{IONTENP} model indicates that the thermal X-ray emission originates from the ISM heated by the forward shock with 
$V_{\mathrm{sh}} \sim 800\,\mathrm{km\,s^{-1}}$, with an ionization timescale of $\sim 10^{11}\,\mathrm{cm^{-3}\,s}$.
The best-fit shock velocity is consistent with the previous optical study \citep{morse1996}.
The best-fit $\tau \sim 10^{11}\,\mathrm{cm^{-3}\,s}$ is also consistent with a product of the postshock electron density of $\sim 12\,\mathrm{cm^{-3}}$ \citep{morse1996} and the flow time of $\sim 10^{10}\,\mathrm{s}$, which is derived from the width devided by the shock velocity assuming the distance of 50 kpc to N132D.
Our model thus successfully estimates the shock velocity based on the thermal X-ray emission from shock-heated plasma, i.e., the thermal energy provided by the shock wave.

\begin{figure}[t]
    \centering
    \includegraphics[width=8.5cm]{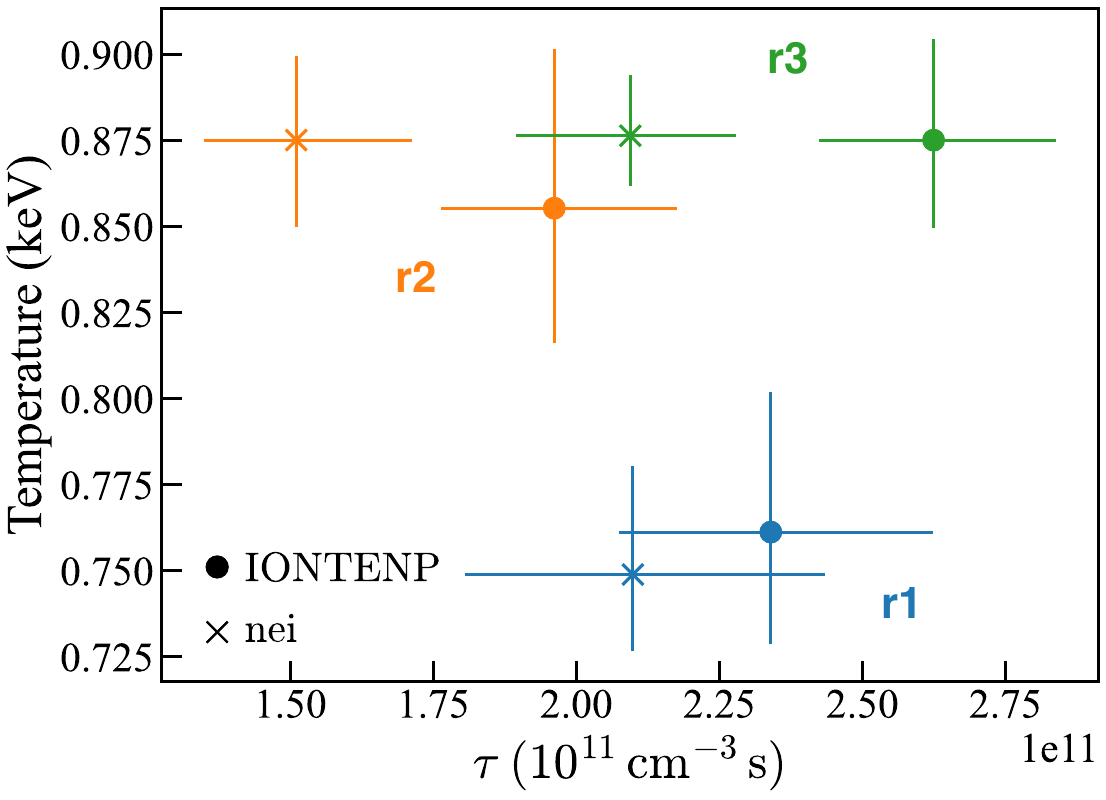}
    \caption
    {
        The best-fit pairs of the electron temperature and $\tau$ determined with the \texttt{IONTENP} (filled circles) and \texttt{nei} (crosses) in each source region.
    }
    \label{fig: ComparisonBestfitParams}
\end{figure}

We compare the best-fit parameters of the \texttt{IONTENP} model with those of the \texttt{nei} model.
Since the electron temperature of the \texttt{IONTENP} model is a function of $V_{\mathrm{sh}}$, $\tau$, and $\beta$, a direct comparison between the two models can be made. 
The results are shown in Figure~\ref{fig: ComparisonBestfitParams}, where the best-fit values of the \texttt{IONTENP} model are converted to the electron temperatures.
We find that, for each region, the electron temperatures of the two models are consistent with each other, but the $\tau$ value is systematically lower in the \texttt{nei} model than in the \texttt{IONTENP} model, which is consistent with the discussion in Section~\ref{subsec: systematic bias}.

For future prospects, the \texttt{IONTENP} model could be updated to account for the supply of elements through the dust destruction process, which is likely efficient in N132D \citep{williams2006,dopita2018}.
Efficient dust destruction predicts a lower $\tau$ value for refractory elements (e.g., silicon and iron) compared to volatile elements (e.g., nitrogen and oxygen), which could be distinguished using high-resolution X-ray spectroscopy.

\section{Conclusions} \label{sec: Conclusions}
In this study, we developed a self-consistent model for thermal X-ray emission from shock-heated plasmas that accounts for both temperature and ionization non-equilibrium conditions.
We considered two types of plasmas with different compositions: solar abundance and pure-iron plasmas.
Our calculations showed that the ionization cooling effect is particularly significant for pure-iron plasmas.
We synthesized the thermal X-ray spectra by combining our model calculations with the AtomDB spectral code.
Comparison of the resulting spectra with those of the constant-temperature equilibrium ionization (\texttt{nei}) model revealed a 30\% underestimation of the ionization degree.
Furthermore, we found that the timescale for ionization equilibrium among ions is shorter than that between ions and electrons, which is crucial for interpreting line widths, as the contribution of the thermal motion of ions diminishes as they approach equilibrium.
We implemented our model in XSPEC and applied it to the archival Chandra X-ray dataset of the SNR N132D, which provided a constraint on the shock velocity of approximately $800~\mathrm{km\,s^{-1}}$, in agreement with previous optical studies.
This model offers a useful tool for measuring shock velocities in supernova remnants from their thermal X-ray emissions, independent of proper motion measurements.
It might also be applicable to other sudden heating events such as solar flares.
The XSPEC-readable model files used in this work are available on our website (TBA).

\begin{acknowledgments}
    The authors thank the referee, John C. Raymond, for his constructive comments and suggestions.
    This work was supported by Grants-in-Aid for Scientific Research (KAKENHI) of the Japanese Society for the Promotion of Science (JSPS) grant Nos.\ JP22KJ1047 (Y.O.), JP22KJ3059 (H.S.), JP24K17093 (H.S.), JP22H00158 (H.Y.), and JP23H01211 (H.Y.).
\end{acknowledgments}

%

\vspace{5mm}
\facilities{Chandra (ACIS)}


\software{
    HEAsoft \citep{nasahighenergyastrophysicssciencearchiveresearchcenterheasarc2014},
    CIAO \citep{fruscione2006},
    astropy \citep{astropycollaboration2013, astropycollaboration2018, astropycollaboration2022},
    PyAtomDB \citep{foster2020}
}



\appendix

\section{Observation and Data Reprocess}
\label{appendix: observation}
The SNR N132D was observed by Chandra with 28 separate sequences (with different satellite roll angles) in the period between March 2019 and July 2020 using the Advanced CCD Imaging Spectrometer (ACIS) S3 chip.
The data reprocessing was performed following the standard procedure of the task \texttt{chandra\_repro} in the Chandra Interactive Analysis of Observations (CIAO) software version 4.14, with the latest Chandra Calibration Database (v.4.9.8).
After the reprocessing, we merged all the data using the CIAO tool \texttt{reproject\_obs} and obtained a total effective exposure of 867 ks.

\section{Background Spectrum Modeling}
\label{appendix: background}

\begin{figure}[t]
    \centering
    \includegraphics[width=8.5cm]{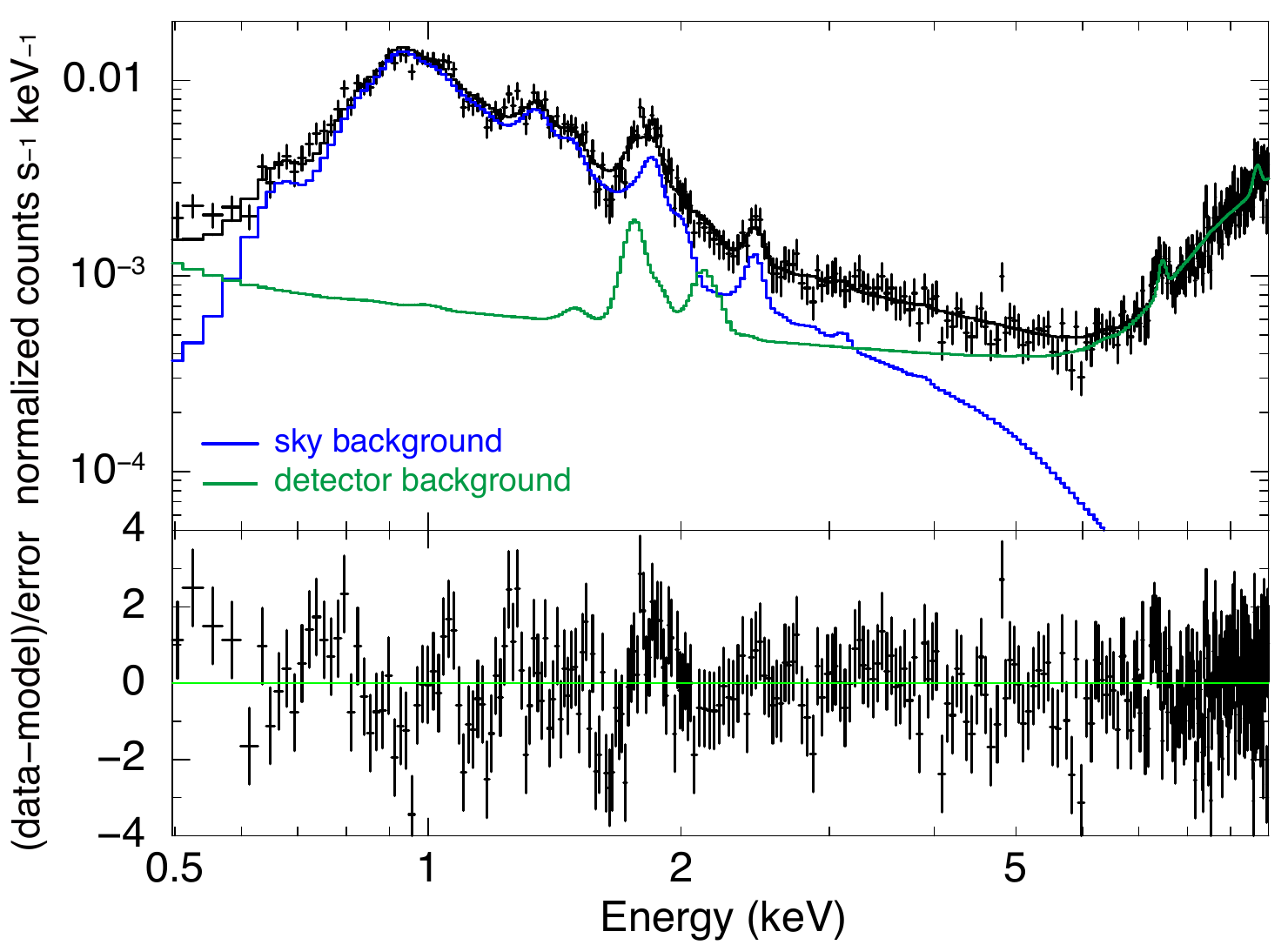}
    \caption
    {
        Chandra ACIS-S background spectrum of N132D (black crosses).
        The best-fit models for sky and detector background components are shown in blue and green lines, respectively.
    }
    \label{fig: background_spectrum}
\end{figure}

\begin{table}[t]
    \caption{Best-fit parameters for the background spectrum of N132D.}
    \label{Tab: best-fit of background}
    \centering
    \begin{tabular}{llc}
        \hline\hline
        Model & Parameter & Value \\
        \hline
        \texttt{TBabs} & $N_{\rm H}$ ($10^{22}$ cm$^{-2}$) & $0.34^{+0.03}_{-0.03}$ \\
        \texttt{apec}  &  $kT$ (keV)  & $0.22^{+0.01}_{-0.01}$ \\
        & normalization\tablenotemark{a} & $3.1^{+0.9}_{-0.7} \times 10^{-4}$  \\
        \texttt{apec} & $kT$ (keV) & $0.90^{+0.03}_{-0.03}$ \\
        & normalization\tablenotemark{a} & $4.5^{+0.3}_{-0.3} \times 10^{-5}$ \\
        \texttt{powerlaw} & Photon index & 1.4 (fixed) \\
        & Normalization\tablenotemark{b} & $4.8^{+0.4}_{-0.4} \times 10^{-6}$ \\
        \hline
        C-stat/d.o.f. &  & 702.62/650 \\
        \hline
    \end{tabular}
    \tablenotetext{a}{The normalization of the \texttt{apec} model is defined as $\frac{10^{-14}}{4 \pi D^2} \int n_{\mathrm{e}} n_{\mathrm{H}} \mathrm{d}V$, where $D$ is the distance to the source, $n_{\mathrm{e}}$ is the electron number density, and $n_{\mathrm{H}}$ hydrogen number density.}
    \tablenotemark{b}{Normalization of the power-law model in units of cm$^{-2}$ s$^{-1}$ keV$^{-1}$ at 1 keV.}
\end{table}

We model the background spectrum instead of subtracting it from the source spectrum.
The background spectrum consists of two components: the detector background and the sky background.
The detector background is modeled using the tool \texttt{mkacispback} \citep{suzuki2021}.
This task estimates the particle-induced background component for an input region.
The generated model by \texttt{mkacispback} allows the entire normalization to be changed as a free parameter.
The sky background includes contributions from the local hot bubble, the Milky Way halo \citep{kuntz2000}, and the cosmic X-ray background \citep{chen1997}.
Since our goal is to obtain an empirical model of the background spectrum, we employ an approximated model consisting of two optically thin thermal plasma emissions and a power low with foreground absorption; \texttt{TBabs*(apec+apec+powerlaw)}.
The foreground absorption is the Tuebingen-Boulder interstellar medium absorption (\texttt{TBabs}) model \citep{Wilms2000} with the hydrogen column density as a free parameter.
The thermal emission components (\texttt{apec}) represent emissions from the Milky Way halo and the hot gaseous halo \citep{nakashima2018,kavanagh2020}.
For both \texttt{apec} models, electron temperature $(kT_{\mathrm{e}})$ and normalization are treated as free parameters.
The index of power low is fixed at 1.46 \citep{chen1997} and the normalization is free to vary.
We fit the background spectrum in the 0.5--10.0 keV band.
The fit results are shown in Figure~\ref{fig: background_spectrum}.
The obtained best-fit parameters are summarized in Table~\ref{Tab: best-fit of background}.



\end{document}